\documentclass[aps,prapplied,reprint]{revtex4-2}

\usepackage[utf8]{inputenc}
\usepackage{amsmath}
\usepackage{amssymb}
\usepackage{amsfonts}
\usepackage{graphicx}
\usepackage{siunitx}					% mise en forme des unités
\usepackage{color}					% gestion de différentes couleurs

\usepackage[colorlinks=true,citecolor=blue,urlcolor=red]{hyperref}

\begin{document}

\title{Harmonic calibration of quadrature phase interferometry}

\author{Baptiste Ferrero}
\affiliation{Univ Lyon, ENS de Lyon, CNRS, Laboratoire de Physique, F-69342 Lyon, France}

\author{Ludovic Bellon}
\email{ludovic.bellon@ens-lyon.fr}
\affiliation{Univ Lyon, ENS de Lyon, CNRS, Laboratoire de Physique, F-69342 Lyon, France}

\date{\today}

\begin{abstract}
The two output signals of quadrature phase interferometers allow to benefit both from the high sensitivity of interferometry (working inside a fringe) and from an extended input range (counting fringes). Their calibration to reach a linear output is traditionally performed using Heydemann's correction, which involves fitting one output versus the other by an ellipse. Here we present two alternative methods based on the linear response of the measurement to a sinusoidal input in time, which enables a direct calibration with an excellent linearity. A ten fold improvement with respect to the usual technique is demonstrated on an optical interferometer measuring the deflection of scanning force microscopy cantilevers.
\end{abstract}

\maketitle 

Interferometers are todays gold standard when it comes to measuring displacements or deformations with a high precision\cite{Schodel-2021}. Gravitational wave interferometers\cite{Allocca-2020}, as glaring examples, reach an impressive resolution down to $\SI{e-20}{m/\sqrt{Hz}}$. Such an achievement is obtained by maintaining the working point around the maximum sensitivity of the detector, thus forbidding simple measurements of large deformations $d$. Indeed, the output of interferometers is always non-linear with their input $d$: the optical power $I$ at their output is a periodic function of an optical phase $\varphi \propto d/\lambda$, where $\lambda$ is the wavelength of the source. The linearization of the output is then only possible for a fraction of the wavelength: $d\ll\lambda$. In the simplest case (single reflection of the probe beam on the moving part), this periodic function is sinusoidal:
\begin{equation} \label{eqI}
I=I_0+I_1\cos (\varphi),
\end{equation}
with $\varphi = 4\pi d/\lambda$, and $I_0$ and $I_1$ two parameters that can be calibrated by exploring a range in $d$ larger than $\lambda/2$, thus a range in $\varphi$ larger than $2\pi$. An adequate workaround to circumvent the non-linear output is then to create within the interferometer a second optical signal in quadrature with the first one\cite{Schodel-2021,Heydemann-1981}:
\begin{equation} \label{eqQ}
Q=Q_0+Q_1\sin (\varphi + \psi),
\end{equation}
with $Q_0$, $Q_1$, and $\psi$ (respectively offset, amplitude and deviation to perfect quadrature) three more parameters that can be calibrated with a large excursion in $\varphi$.

\begin{figure}[htbp]
\begin{center}
\includegraphics{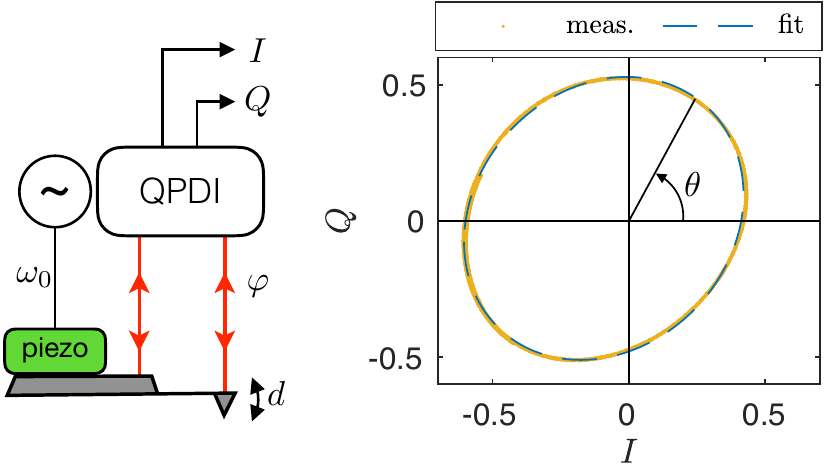}
\caption{We use a quadrature phase differential interferometer (QPDI) to measure the deflection of an atomic force microscopy (AFM) cantilever\cite{Paolino-2013}.  The optical phase $\varphi = 4 \pi d/\lambda$ between the two laser beams after reflection on the support and free end of the cantilever encodes the deflection $d$ of the latter, with an absolute calibration with respect to the wavelength $\lambda=\SI{633}{nm}$ of the laser. The QPDI produces two outputs in quadrature, $I$ and $Q$. Imposing an harmonic oscillation of $d$ at angular frequency $\omega_0$ through a piezoceramic, we explore the full $[0,2\pi]$ range of phases. Plotting $Q$ versus $I$ gives the calibration curve, where lies all measurements, and that can be parametrised by the polar angle $\theta$. Calibrating the interferometer means finding the correspondance between $\theta$ and $\varphi$. In Heydemann's correction scheme\cite{Heydemann-1981}, this is realised by fitting the calibration curve by an ellipse.}
\label{QPDI+ellipse}
\end{center}
\end{figure}

In practice in a $Q$ versus $I$ plot, we expect data to lay on an ellipse from eqs.~\eqref{eqI} and \eqref{eqQ}. The ideal case, $I_1=Q_1$ and $\psi=0$, leads to a circle: the polar angle $\theta$ of the measurement point on this circle is then a direct measurement of the optical phase $\varphi$. To account for imperfections of the actual instrument, it is customary to use Heydemann's correction\cite{Heydemann-1981}: the closed curve in the $I,Q$ plane corresponding to a large enough excursion in $\varphi$ (the calibration curve) is fitted by an ellipse to extract the five parameters $I_0$, $I_1$, $Q_0$, $Q_1$ and $\psi$, as illustrated in Fig.~\ref{QPDI+ellipse}. Eqs.~\eqref{eqI} and \eqref{eqQ} are then reversed to compute the optical phase:
\begin{subequations}
\begin{align}
\cos \varphi &= \frac{I-I_0}{I_1} \\
\sin \varphi &=  \frac{Q-Q_0}{Q_1\cos \psi} + \frac{I-I_0}{I_1} \tan \psi.
\end{align}
\end{subequations}
Through the measurements of $I$ and $Q$ and the knowledge of the five calibration parameters, the simultaneous knowledge of $\cos \varphi$ and $\sin \varphi$ allows to extract of $\varphi$ in the full $[0, 2\pi]$ interval from their signs and the arctan function applied to their ratio. Unwrapping $\varphi$ as time runs can then be used to reach a virtually infinite input range, while maintaining the full sensitivity of the interferometric measurement for sub-fringe variations. This approach, first introduced by Heydemann\cite{Heydemann-1981}, has been used in many devices and experiments, to linearise the dual output of quadrature phase interferometers with a very good precision\cite{Wu-1996,Petru-1999,Eom-2001,Bellon-2002,Povar-2011,Paolino-2013,Bridges-2021}.

Though this approach can handle many imperfections of the interferometer, such as offsets, $I$ and $Q$ imbalance, imperfect quadrature, it still rely on the hypothesis that the periodic outputs are simple sinusoidal functions of the optical phase $\varphi$. Other imperfections, such has beam clipping, can actually create more complex periodic functions, than manifest as a calibration curve in the $I,Q$ plane that deviates from an ellipse, and higher order terms in the Fourier expansion of $I(\varphi)$ and $Q(\varphi)$. One could add such terms a priori, and fit the $I,Q$ curve with a more general parametric curve, to extract more calibration parameters. The number of additional fitting terms to include would however need to be guessed, and the inversion problem might be difficult to perform.

In this Letter, we use an alternative approach to calibration, which is in some sense more conventional. Let us first describe the $I,Q$ plane by its complex number representation $z=I+iQ-z_0$, with $i=\sqrt{-1}$ the imaginary number and $z_0$ the origin. We place $z_0$ somewhere inside the calibration curve (for exemple at midway between the minimum and maximum of $I$ and $Q$). Any measurement point can now be written as $z=|z|e^{i\theta}$, with $\theta$ the argument of $z$ (or the polar angle) in this representation. For simple calibration curves (close to a circle or an ellipse for exemple), the relation between $\theta$ and $\varphi$ is bijective, and its knowledge is a calibration: once this bijection $\theta=\Theta(\varphi)$ is known, any measurement point $z$ directly leads to the optical phase: $\varphi=\Theta^{-1}(\arg z)$. It would be convenient to be able to apply a known and calibrated deformation $d$, thus optical phase $\varphi$, spanning the full $[0, 2\pi]$ interval, and directly plot the measured $\theta$ versus the applied $\varphi$ to access the calibration function $\Theta$. However, such an approach supposes that we already have a calibrated instrument as precise as the interferometer we want to calibrate...

The trick we implement in our approach is to use a pure harmonic calibration signal, obtained with a driving at a single frequency of a resonant system. Our setup, described in Refs. \onlinecite{Bellon-2002, Paolino-2013} and Fig.~\ref{QPDI+ellipse}, uses a quadrature phase interferometer to measure the deflection of a micro-cantilever used in an atomic force microscope (AFM). Using a waveform generator with a very low distortion, we drive the cantilever at its resonance frequency through a piezo-ceramic. The amplitude of the driving is small enough to have a linear response of the electro-mechanical components, all the more as the resonant behavior of the cantilever with a high quality factor efficiently filters out any signal at frequency higher than the resonance. The deflection $d$ of the cantilever can thus be considered purely harmonic at angular frequency $\omega_0$, so that the optical phase writes: $\varphi(t)=\varphi_0+\varphi_1 \cos[\omega_0 (t-t_0)]$. $\omega_0$ is set by the operator, so if we manage to identify the three parameters $\varphi_0$, $\varphi_1$ and $t_0$, we will be able to plot $\varphi(t)$ (imposed) versus $\theta(t)$ (measured), and have a direct access of the calibration function $\Theta^{-1}$.

\begin{figure}[tbp]
\begin{center}
\includegraphics{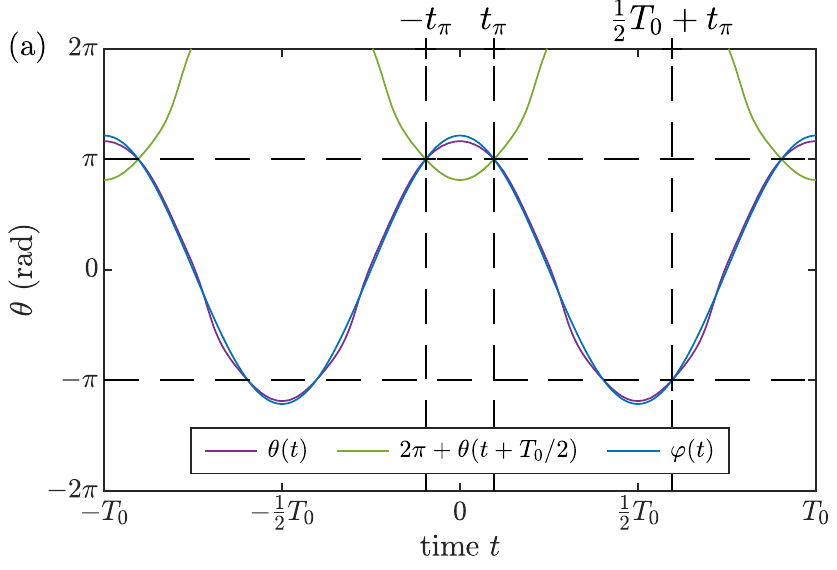}

\vspace{2mm}

\includegraphics{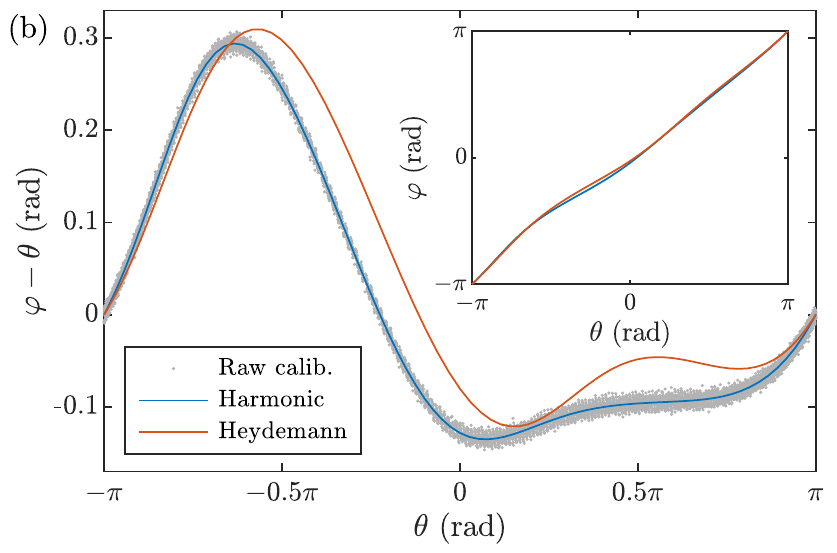}
\caption{(a) Time trace $\theta(t)$ superposed with $2\pi+\theta(t+\frac{1}{2}T_0)$ during calibration, with $T_0$ the driving period. For periodicity reasons, the intersection of those curves define the time $t_\pi$ where $\varphi(\pm t_\pi)=\pi$, entirely defining the optical phase $\varphi(t)$. (b) Since $\varphi(t)$ versus $\theta(t)$ is very close to the identity (inset), we plot the correction $\varphi-\theta$ to add to the polar angle $\theta$ to compute the optical phase $\varphi$. Heydemann's correction (red line), interpreting the data from the ellipse fitted on the calibration curve, and the harmonic calibration (blue line), proposed in this Letter, yield similar results, with a deviation up to $\SI{0.3}{rad}$ from the identity. The harmonic calibration function is computed with a low pass filtering of the experimental data (gray dots).}
\label{Figthetaphi}
\end{center}
\end{figure}

$\varphi_0$ is usually arbitrary in interferometry, and correspond to the origin of phase (or deflection in our case), we thus impose $\varphi_0=0$ without loss of generality. Let us continue with $t_0$: since $\varphi$ is maximum at $t=t_0$ (and at every period $T_0=2\pi/\omega_0$), so is $\theta$ : $\dot{\theta}(t_0)=\Theta'[\varphi(t_0)]\dot{\varphi}(t_0)=0$.  We can therefore define the origin of time at a maximum of the measured $\theta$ to set $t_0=0$. We are therefore only left with
\begin{equation} \label{eqphit}
\varphi(t)=\varphi_1 \cos(\omega_0 t),
\end{equation}
and we just need to extract the value of $\varphi_1$ from the measurement. Let us first note that since we explore with the closed calibration curve the full range $[0, 2\pi]$, we necessarily have $\varphi_1>\pi$. Let us then define the smallest positive time $t_\pi$ when $\varphi(t_\pi)=\pi$. From eq.~\eqref{eqphit}, we know that half a period later, we will have $\varphi(t_\pi+\frac{1}{2}T_0)=-\pi=\varphi(t_\pi)-2\pi$. Since both phases are equal modulo $2\pi$, the measurement point is the same on the calibration curve. Unwrapping $\theta$ as well, we should have $\theta(t_\pi+\frac{1}{2}T_0)=\theta(t_\pi)-2\pi$. When plotting $\theta(t)$ and $2\pi+\theta(t+\frac{1}{2}T_0)$, we thus see the two curves intersecting in $t=t_\pi$, as illustrated in Fig.~\ref{Figthetaphi}(a). By symmetry, they should also intersect in $t=-t_\pi$. Note that the time origin can also be defined to meet this symmetry, instead of looking for a maximum of $\theta$. Once $t_\pi$ is graphically determined, we directly have the value of $\varphi_1$, thus the full knowledge of $\varphi(t)$ with eq.~\eqref{eqphit}: since $\varphi(t_\pi)=\pi$, then $\varphi_1=\pi/\cos(\omega_0 t_\pi)$, and  
\begin{equation}
\varphi(t)=\pi \frac{\cos(\omega_0 t)}{\cos(\omega_0 t_\pi)}.
\end{equation}
Finally, plotting $\varphi(t)$ versus the measured $\theta(t)$ directly leads to the calibration function $\Theta^{-1}$, as illustrated in Fig.~\ref{Figthetaphi}(b). To smooth the experimental raw data, we take advantage of the $2\pi$ periodicity in $\theta$  of $\varphi-\theta$: first, we average the experimental data on 512 evenly distributed points in the $[-\pi,\pi]$ interval, then we remove high frequency noise by low pass filtering of this curve (Fast Fourier Transform (FFT) of the curve, remove high frequency and low amplitude components, inverse FFT)\cite{Ferrero-2022-Dataset}. The resulting curve can be interpolated to any input data $\theta$ to compute the optical phase.

\begin{figure}[b]
\begin{center}
\includegraphics{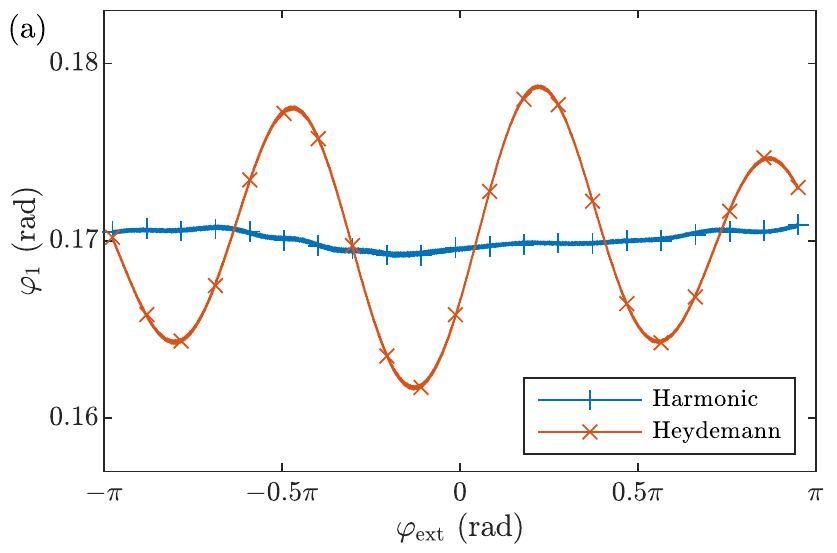}

\vspace{2mm}

\includegraphics{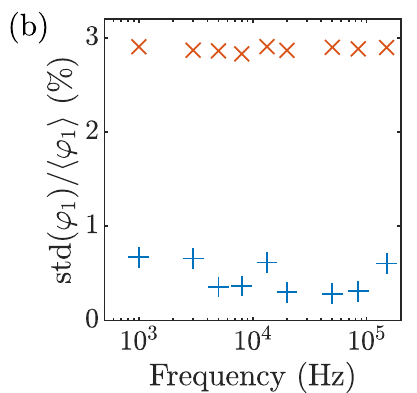}\includegraphics{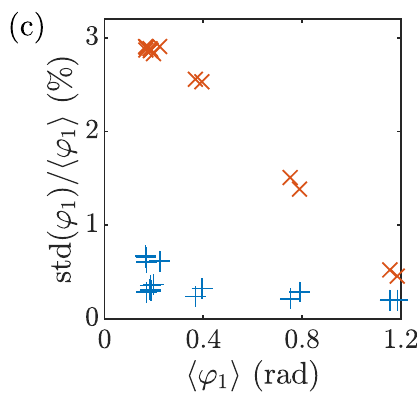}
\caption{(a) We apply steady driving to the cantilever at $\SI{50}{kHz}$ while slowly changing the mean working point with an external optical phase $\varphi_\mathrm{ext}$. The oscillation amplitude $\varphi_1$ at $\SI{50}{kHz}$ extracted from Heydemann's correction ({\color{red}$\times$}) shows a $\SI{\pm3}{\%}$ rms variation around the mean value, whereas it is almost constant when using the harmonic calibration ({\color{blue}$+$}). We use the normalised standard deviation $\mathrm{std}(\varphi_1)/\langle \varphi_1\rangle$ as a figure of merit for the linearity of the measurement, and report it as a function the oscillation frequency (b, for an amplitude around $\SI{0.2}{rad}$) or amplitude (c, amplitudes above $\SI{0.3}{rad}$ only for the first and second resonant modes of the cantilever, at $f_0=\SI{13.35}{kHz}$ and $f_1=\SI{84.64}{kHz}$).}
\label{FigLinTest}
\end{center}
\end{figure}

To illustrate the gain in linearity from this calibration process with respect to the Heydemann correction, we perform the following experiment: after performing the calibration step, we apply to the cantilever a steady harmonic driving of low amplitude (deflection $\SI{10}{nm}$ to $\SI{60}{nm}$, corresponding to an optical phase amplitude $\varphi_1=\SI{0.2}{rad}$ to $\SI{1.2}{rad}$) and high frequency (from $\SI{1}{kHz}$ to $\SI{150}{kHz}$), and at the same time we artificially induce a low frequency drift (around $\SI{1}{Hz}$) of the working point, to explore the full calibration curve. In our setup, this drift is controlled optically by adding an external phase $\varphi_\mathrm{ext}$ to the optical one $\varphi$, it could also be done by adding a low frequency force of large amplitude applied to the cantilever tip. The high frequency deflection amplitude, measured with a digital lock-in data processing, should remain constant, regardless of the working point, and thus probes the linearity of the output in the full input range. We report an example of this procedure in Fig.~\ref{FigLinTest}(a), for a $\sim\SI{0.2}{rad}$ oscillation at $\SI{50}{kHz}$. The amplitude extracted from Heydemann's correction shows a $\SI{\pm3}{\%}$ rms variation around the mean value while varying $\varphi_\mathrm{ext}$, whereas the harmonic calibration yields a amplitude constant within $\SI{\pm0.3}{\%}$. This one order of magnitude gain in linearity is true at any probed frequency, as show in Fig.~\ref{FigLinTest}(b). When probing larger amplitude, the non-linearity is averaged out in Heydemann's approach as the oscillation averages the sensitivity on a larger range, as shown by Fig.~\ref{FigLinTest}(c). In the meantime, it stays equally good using the harmonic calibration.

\begin{figure}[tbp]
\begin{center}
\includegraphics{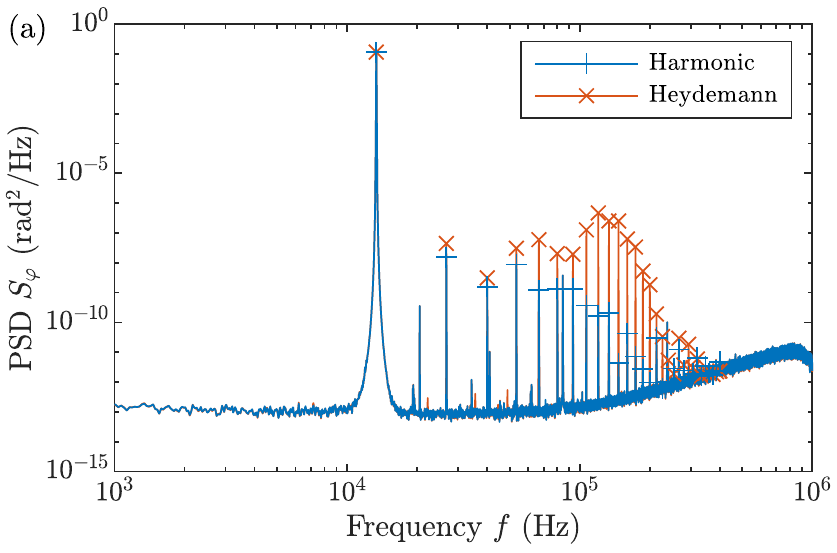}

\vspace{2mm}

\includegraphics{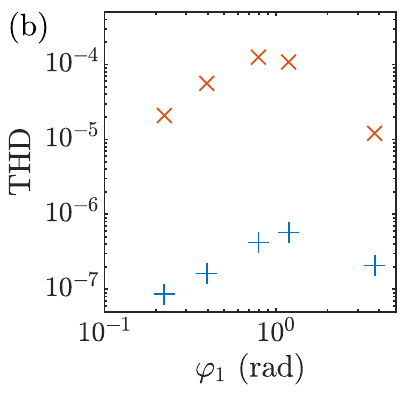}\includegraphics{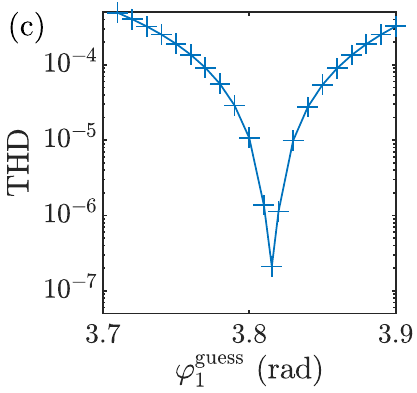}
\caption{(a) Power Spectrum Density (PSD) $S_\varphi$ of the optical phase $\varphi$ while driving to the cantilever at its first resonance frequency $f_0=\SI{13.35}{kHz}$ with an amplitude $\varphi_1=\SI{3.8}{rad}$. All harmonics of this fundamental frequency (markers) are created by the residual non-linearities of the detection scheme. Heydemann's correction produces higher distortion than the harmonic calibration for all harmonics. (b) Total Harmonic Distortion (THD, ratio of all harmonics power over the fundamental) versus oscillation amplitude with a driving at $f_0$: the harmonic calibration is consistently better than the classic approach. (c) THD versus $\varphi_1^\mathrm{guess}$ for the calibration dataset: if we let the amplitude $\varphi_1^\mathrm{guess}$ as a adjustable parameter and compute for each guess the calibration function $\Theta^{-1}$, then the THD of $\Theta^{-1}\big(\theta(t)\big)$, the latter presents a sharp minimum for the actual value of $\varphi_1$, here $\SI{3.815}{rad}$.}
\label{FigPSD}
\end{center}
\end{figure}

Another way to probe the non-linearity of the output is to drive the cantilever at a given frequency and look for the generation of harmonics by the measurement process. We plot in Fig.~\ref{FigPSD}(a) the Power Spectrum Density (PSD) of the deflection during the calibration run: the cantilever is driven at its first resonance $f_0=\SI{13.35}{kHz}$ with an amplitude $\varphi_1=\SI{3.815}{rad}$ [data corresponding to Fig.~\ref{Figthetaphi}(a)]. We can observe how the peaks corresponding to the harmonics at integer multiples of $f_0$ decreases when choosing the harmonic calibration over Heydemann's correction.  To be quantitative, we can compute the Total Harmonic Distortion (THD), defined as the integral of the PSD around all harmonics of the excitation frequency, normalised by the one around $f_0$. For this specific amplitude and frequency, we measure a THD of $\num{2e-7}$ for the harmonic calibration, down from $\num{e-5}$ for the classic approach. This two orders of magnitude gain in THD is consistent on every oscillation amplitude explored, as shown in Fig.~\ref{FigPSD}(b).

The THD can be used as an indicator of the performance of the calibration, but it can also be used to perform the calibration itself. Indeed, as mentioned earlier, the important parameter to extract from the calibration dataset is the imposed oscillation amplitude $\varphi_1$, which can be done using the periodicity trick of Fig.~\ref{Figthetaphi}(a). However, one can directly make a guess $\varphi_1^\mathrm{guess}$ for $\varphi_1$, compute a calibration function $\Theta^{-1}$, and then the THD of the reconstructed phase $\Theta^{-1}\big(\theta(t)\big)$ from the same calibration dataset. We then expect non-linearities to be present in the signal if the guess value $\varphi_1^\mathrm{guess}$ differs from the actual value $\varphi_1$, thus degrading the THD. As shown in Fig.~\ref{FigPSD}(c) by plotting the THD versus $\varphi_1^\mathrm{guess}$, the curve presents a sharp minimum at $\varphi_1$. This minimisation procedure is another way to perform the harmonic calibration, and the relative difference in $\varphi_1$ computed from both methods (periodicity trick or THD minimum) is only $\num{e-4}$ with the dataset of Fig.~\ref{Figthetaphi}(a): both calibrations lead to indistinguishable results. The benefits of the THD minimisation approach is that it is very general : it doesn't require the periodicity of the output, nor the minimum amplitude $\varphi_1>\pi$. As a matter of fact, it could be applied to any measurement device having a non-linear output, for which applying a pure harmonic input is possible.

To summarize, in this Letter we present a generic method to calibrate the dual output of quadrature phase interferometers, based on applying a pure harmonic input signal. Using a large enough oscillation (interferometer phase spanning more than $2\pi$), and the periodicity properties of the input and outputs, we can infer the amplitude of the input with no need for any other information. Plotting the inferred input versus the measured output yields the calibration function, which can be subsequently be used for any input signal. We demonstrate on an experiment using a differential interferometer measuring the deflection of an AFM cantilever that a significant gain in linearity can be achieved with respect to the common Heydemann's correction approach. The THD can be used as a figure of merit for the calibration process, and can actually take part to the calibration procedure itself: its minimisation from guess values of the input amplitude bypasses the periodicity trick.

Undoubtedly, the gain in performance is setup (and user) dependant: a perfectly tuned interferometer, with negligible imperfections, wouldn't benefit much from such correction (just as Heydemann's wouldn't be necessary in such case). However, in real life use, the approach is light to implement and could benefit many existing interferometric devices. If signal post-processing is sufficient, it can be performed with simple data analysis software. Once the calibration curve extracted, a real-time implementation could also be performed using FPGA based devices.

To conclude this Letter, let us mention two interesting perspectives for this work. First, multi-frequency AFM\cite{Garcia-2012}: in this approach, one uses the non-linear characteristic of the tip-sample interaction to create from a single (at most a few) frequency driving of the cantilever a comb of response frequencies. The amplitude of those harmonics can then be used for imaging, or even to reconstruct the non-linear interaction potential\cite{Forchheimer-2012}. The linearity of the detection is crucial in this technique, as one would like harmonics to come from the physics of the interaction rather than from the sensor. Moreover, the signal to noise ratio is important to get as many harmonics as possible: the more the merrier when its comes to the inversion problem. On those 2 criteria, an interferometric readout of the cantilever deflection (as the one described in this Letter) with high sensitivity and linearity would be beneficial. A second interesting perceptive is the very general calibration procedure offered by the THD minimisation approach. Indeed, it could be applied to any measurement device having a non-linear output for which applying a pure harmonic input is possible. In such case, one can use guess amplitudes for the input, create a calibration curve, infer the THD, and then minimize the latter with respect to the guess value. As long as the instrument bandwidth is large enough to include a few harmonics of the forcing, this procedure can be applied and yields a direct linearization of the device output. Interestingly, the non-linear character of the output is instrumental for this linearization approach to work !

\medskip 
\noindent \textbf{Acknowledgments} This work has been financially supported by the French r\'egion Auvergne Rh\^one Alpes through project Dylofipo and the Optolyse plateform (CPER2016). We thank S. Ciliberto, C. Crauste, R. Pedurand, A. Labuda for enlightening scientific and technical discussions.

\medskip
\noindent \textbf{Data availability} The datasets and data processing tools to compute calibration functions that support the findings of this study are openly available in Zenodo at \url{https://doi.org/10.5281/zenodo.6801118}~\cite{Ferrero-2022-Dataset} 

\bibliography{HarmonicCalib}

\end{document}